\input{epsf}

\documentclass[preprint,showpacs,preprintnumbers,amsmath,amssymb]{revtex4}

\usepackage{graphicx}% Include figure files

\begin{document}
\draft

\title{Could a reported 2007 analysis of Super-Kamiokande data have missed a detectable supernova signal from Andromeda?}

\author{Robert Ehrlich}
\address{George Mason University}
\date{\today}
\begin{abstract}
According to a 2007 paper there was no evidence for a neutrino burst of two or more events in Super-Kamiokande (SK) during the entire period of data-taking from 1996 to 2005 from Andromeda or anywhere else.  There is, however a scenario under which a detectable signal could have been missed given the search method employed by the analysis, and it would have been found using an alternate method.  The alternate method depends on the hypothesis that two of the neutrino mass eigenstates have masses 4.0 eV and 21.4 eV which was inferred from an analysis of the SN 1987A data.  Although one might argue that the hypothesis of such large neutrino masses is remote, there is a way they could be compatible with observed constraints on neutrino masses involving a third tachyonic ($m^2<0$) eigenstate, plus three sterile neutrinos.  Given the importance of a positive supernova search result, and the ease of conducting it using existing SK data, there would seem to be little reason not to do it.  
\end{abstract}


\maketitle
\section{Introduction}

To date the only supernova giving rise to an observed neutrino burst is SN 1987A for which 24 events were seen in 3 detectors in a burst lasting about 15 s \citep{Kamioka, IMB, Baksan}.  Given the probable significant delay before future Mton neutrino detectors are completed, there is considerable interest in maximizing the efficiency for supernova detection using the best detectors currently available, especially the 50 kton Super-Kamiokande (SK).  Two important ways of enhancing the sensitivity involve modifying SK in some way, or alternatively selecting an optimum energy threshold cut $E_{th}$ to impose on the data that is based on the very different dependencies of the expected signal and background events \citep{Vagins, Scholberg}.  Of course, a supernova from a core collapse within our galaxy would be completely obvious in SK without needing to maximize its detection efficiency, and one well beyond Andromeda likely would not yield a single event in current detectors.  This makes the possibility of seeing a supernova burst from Andromeda (expected signal 2 events in SK) one of great interest.  Thus, given the spatial distribution of distances of potential sources, Andromeda plays a very special role in extragalactic supernova searches with SK.  In fact, since the expected frequency of supernovae closer than a distance d does not increase significantly between the edge of our galaxy and Andromeda, and also between Andromeda (d = 778 kpc) and the next rich source at a distance of 2.7 Mpc, at which point the expected flux would have dropped by a further factor of 10 by virtue of the inverse square law \citep{Ando}.    One may therefore conclude that if one should observe genuine signal of 2 - 3 events in SK the source is very likely to be Andromeda -- a fact that could be checked based on directional information on those events.

A 2007 paper analyzed data from SK for the period 1996 - 2005, and it reported no signal from Andromeda or anywhere else \citep{Ikeda}.  However, for the pre-2002 period SK had uncertainties in its energy calibration and it was not used in the 2007 analysis.  While the pre-2002 SK could easily show a supernova in our galaxy, it was not sensitive enough to look for a supernova in Andromeda.  Thus, the search for a signal from Andromeda, which required making an energy cut on the data in order to enhance the signal to background ratio, was based only on 3 years of data: 2005 - 2007.  As of 2013 there now exists 7 more years of data to examine for SK, so it might be timely to redo the analysis to look for a low event rate signal from Andromeda.  However, far more important than the additional 7 years of data since 2005 is the possibility that the earlier analysis left open a way that a detectable signal might have escaped notice.

\section{Search method used in 2007 analysis}
Given that the signal from SN 1987A lasted around 15 seconds, the search reported in 2007 used a sliding 20 second time window, and it required the presence of at least N = 2 candidate events (a "cluster") in that window, nearly all of which would be expected to be background.  In order to reduce background a threshold energy cut $E_{th} > 17 MeV,$ was imposed which drastically reduced background to about one event per day, while simultaneously reducing any real supernova signal by about $33\%$, based on the expected supernova spectrum -- thereby reducing the expected counts from a supernova in Andromeda from 2.0 to 1.4 \citep{Scholberg, Ikeda}.  The background was reduced further by considering the spatial positions of event vertices within the detector, since background events tend to cluster spatially, while real neutrino events would not.  Therefore a cut imposed on the average spacing of the vertices for the clusters in any 20 second time window was found to reduce background almost to zero, while having a negligible effect on eliminating any real N event cluster \citep{Ikeda}.  In fact, the combined cuts on energy and vertex proximity left only 3 remaining candidate $N > 1$ event clusters for the entire data-taking period.  Unfortunately, each of those 3 clusters occurred within 0 to 6 seconds after blastings in the mine that housed SK, and they had to be excluded on that basis, thereby reducing the final number of candidates to zero.  

\section{The two mass eigenstate hypothesis}
Most analyses of SN 1987A assume that the 24 events can only set upper limits on the mass of the electron neutrino \citep{Bahcall, Arnett, Pagliaroli, Ellis}.  However, a recent analysis of SN 1987A neutrinos suggests that the (anti)electron neutrinos detected consist of two neutrino mass eigenstates having masses $m_1=4.0\pm 0.5 eV$ and $m_2=21.4 \pm 1.0 eV$ \citep{Ehrlich1, Ehrlich2}, using an idea that was explored much earlier\citep{Huzita, Cowsik}.  Although the suggested masses seem preposterously large, the 2012-2013 papers suggest a way they could be consistent with much smaller empirical upper limits on the neutrino masses \citep{Ehrlich1, Ehrlich2}.  For example, the upper limit on the electron neutrino mass, and the sum of the flavor state masses from cosmology can all be addressed if the third mass eigenstate is tachyonic ($m^2<0$) which would permit all three flavor states to have masses very close to zero.  It is also necessary to have 3 sterile neutrinos nearly degenerate with the three active ones to yield the well established $\Delta m_{atm}^2$ and $\Delta m_{sol}^2$ values.  

The claim of 4.0 eV and 21.4 eV mass eigenstates rests on the idea that most of the 24 neutrinos observed were emitted during a time much less than the 15 second burst, possibly in as short a time as 1 second.  Core collapse modellers differ on the time distribution in neutrino luminosity, and only some of them suggest a very large fraction in the first second.  For example, while Totani et al. show $73\%$ emitted in the first second and $82\%$ in the first 2 s (based on a numerical integration of their Fig. 1),\citep{Totani} other sources suggest significantly less early luminosity \citep{Hudepohl}. Thus, given the uncertainties, one cannot rule out the possibility that a large majority of the observed neutrinos from SN 1987A were emitted during an interval perhaps as short as 1 s.   If one assumes near-simultaneous neutrino emissions then the neutrino arrival times at the detectors depend primarily on their mass $m$ and energy $E$, based on relativistic kinematics according to:

\begin{equation}
\frac{1}{E^2} = \frac{2}{m^2t_0}(t - T)
\end{equation}

where $t$ is the time a given neutrino is detected, $t_0$ is the light travel time from the supernova, and $T$ is some fixed time for all neutrinos associated with the supernova burst, i.e., the time $t=T$ signifies when neutrinos of infinite energy E would have arrived.  Note that equation 1 implies that on a plot of $1/E^2$ versus $t$ all neutrinos having a fixed mass $m$ lie on a straight line of slope $M=2/(m^2t_0)$.  Thus, if neutrinos having two distinct masses were present among the 24 they would all lie on one of two straight lines having slopes $M_1=2/(m_1^2t_0)$ and $M_2=2/(m_2^2t_0)$ that intersect at a point on the $t$ axis.  Remarkably, as shown in \citep{Huzita, Cowsik, Ehrlich1}, this pattern is exactly what the data show (within the uncertainties in $1/E^2$) for every one of the 24 observed neutrinos.  Finally, we note that the goodness of the two straight line fit (from which the claim was inferred) is not significantly impaired if there is a limited spread in emission times.  In a Monte Carlo simulation in which 24 events are generated, assuming an emission time distribution and energy spectrum following Totanti et al.\citep{Totani}, and  assuming the masses of individual neutrinos are either 4.0 eV or 21.4 eV, the agreement of the fit to two straight lines is usually just as good as for the actual data.  Finally, it should be noted that the seemingly implausible claim of two such large mass eigenstates has recently received additional support from a sterile neutrino fit to the dark matter halos in the Milky Way and in galaxy clusters, using  that same pair of masses\citep{Chan}.

\section{The alternative search method based on 2 mass eigenstates}

In the $1/E^2$ versus $t$ plane the search window used in the 2007 search is a rectangle of height $1/E^2=1/17^2 MeV^2$ and width 20 s which slides along the $t$ axis.  This search window is appropriate if we expect a neutrino burst from a source in Andromeda to have a similar width on arrival to that of SN 1987A, i.e., around 15 s.  However, if the arrival times are primarily due to different travel times for neutrinos having different energies and masses, and if the masses are as large as 4.0 eV and 21.4 eV the arriving pulse will broaden the more remote the source, and \emph{individual} neutrino arrival times will correlate with their energy according to Eq.~(1).  Since Andromeda is about 14 times as far as SN 1987A, a 15 s burst from Andromeda would broaden to about $15 \times 14 =210 s$ on reaching Earth.  The alternate search method proposed here, therefore, replaces the fixed width rectangular search window by one consisting of two triangular regions -- see Figure 1 -- which bracket (within expected neutrino energy uncertainties) the two dashed lines having slopes corresponding to the masses 4.0 eV and 21.4 eV.  Thus, the dashed line down the center of the larger triangle has a slope $M=2/(m^2t_0)$ where $t_0 = 2.54 My$ is the light travel time from Andromeda, and where $m = 21.4 eV.$   Similarly, the dashed line bisecting the skinnier triangle has a slope given by the same equation but with $m = 4.0 eV.$  The sides of the 2 triangles are based on a $1\sigma$ $\pm 16\%$ uncertainty in $E$, and hence a $\pm 32\%$ uncertainty in $1/E^2$ \citep{Nakahata}.  As the two-triangle search window slides smoothly along the $t$ axis both triangular regions must of course move together to keep their bottom vertices in contact and on the t-axis.  

Note that unlike the fixed (20 s) width rectangular search window, the width of the two-triangle search window, $\Delta t$ along the time axis is proportional to $1/E_{th}^2,$  so that on a plot like Fig. 1, where the scale of the two axes is $1/E_{th}^2$ and $\Delta t$, it is the rectangular search window that appears to change its width for two different values of $E_{th}.$   For example, a choice $E_{th} = 17 MeV$ yields a width $\Delta t= 63 s$ whereas the choice $E_{th} = 10 MeV$ yields a width $\Delta t= 182 s$ for the two-triangle window.   It is clear that the proposed two-triangle search window might indeed produce candidate events that would be missed by the standard search window -- especially when a threshold $E_{th} = 10 MeV$ were used instead of $E_{th} = 17 MeV$ and the actual area of the search window becomes nine times larger.  As an example, consider the 3 simulated events shown in Figure 1 as black circles and assume that their locations in time are appropriate to the choice $\Delta t= 182 s.$  They clearly are all inside the two-triangle search window with $E_{th} = 10 MeV,$ but only one of the 3 would be picked up as a candidate either for a rectangular search window or for a two-triangle search window with $E_{th} = 17 MeV.$  Hence no $N > 2$ event cluster would register in this case using either of these latter two search windows -- the standard rectangular one or the two-triangle one with too high an energy threshold.

One might argue justifiably that using an energy threshold as low as $E_{th} = 10 MeV$ could generate a number of false signals.  However, while $E_{th} = 17 MeV$ may be the optimum choice in terms of enhancing the signal to background rates, it may not be an optimum choice if very few real events are expected as in the case of a signal from Andromeda.  If we expect only two events in SK from a supernova in Andromeda, then according to Poisson statistics there is a $27\%$ chance of getting 2 events and a $33\%$ chance of getting 3 or more events.  In contrast, if a cut $E_{th} = 17 MeV$ is applied the expected number of events would drop to  1.4 instead of 2.0, and the previous numbers become: a $24\%$ chance of getting 2 events and a $17\%$ chance of getting 3 or more events.   This means that the chances of finding a very persuasive signal ($N > 2$ events) essentially is reduced in half by virtue of the $E_{th} = 17 MeV$ cut.   Furthermore, given that zero events were seen in the original search, the possibility of a few background events masquerading as a signal is less worrisome when one might easily eliminate them if they do not point back to Andromeda.  The authors of the original search did in fact try lowering $E_{th}$ after finding zero events with the original choice, and they still found no signal.  The main argument being advanced in the present paper, however, is not the that a signal could have been missed due to too low an $E_{th} = 17 MeV$ cut, but rather due to the rectangular shape of the search window.

\section{What if a signal is seen using the alternative method?}
The frequency of supernovae in a galaxy is very uncertain, since perhaps as many as $80-90\%$ are hidden by dust \citep{Mattila}.  Andromeda is a galaxy of comparable size to the Milky Way, and the expected frequency of supernovae is also comparable -- perhaps 5 per century, or one every 20 years -- which suggests there could be a 50\% chance that a supernova occurred during observing period without any visible indication of it.  If no signal is found by examining the 2002-2012 record for SK using the alternate search method, it could simply mean that there were no supernovae in Andromeda during the 10 years since 2002.  If 2 to 3 events are observed in SK, one would need to see how well they point back to Andromeda.   At an energy of $10 MeV$ the directional uncertainty of arriving neutrinos is about $12^0 $ \citep{Nakahata}.  A directional uncertainty in the source of about $12^0$ amounts to about $1\%$ of all possible solid angles, so the chances of finding a cluster of N events all pointing back to Andromeda within $12^0$ would be $p = 0.01^N$ which would be highly convincing for $N > 1.$  Were such a result found and the events not picked up within the earlier 20 s wide rectangular search window, it would constitute evidence for a pulse broadening with distance, in conformity with the existence of 4.0 eV and 21.4 eV neutrinos.  Hence, the significance of such a positive result would be far greater than being the first extragalactic observation of a supernova burst of neutrinos, since it would also give support to the hypothesis of two identified mass eigenstates, and all that follows from that hypothesis, including the existence of tachyonic neutrinos.  Such a hypothetical positive search result would constitute a third piece of evidence in support of the two claimed mass eigenstates in addition to the SN 1987A analysis\citep{Ehrlich1, Ehrlich2} and the sterile neutrino dark matter fits.\citep{Chan}.

\newpage

\begin{figure}[h]
\epsfxsize=250pt 
\epsffile{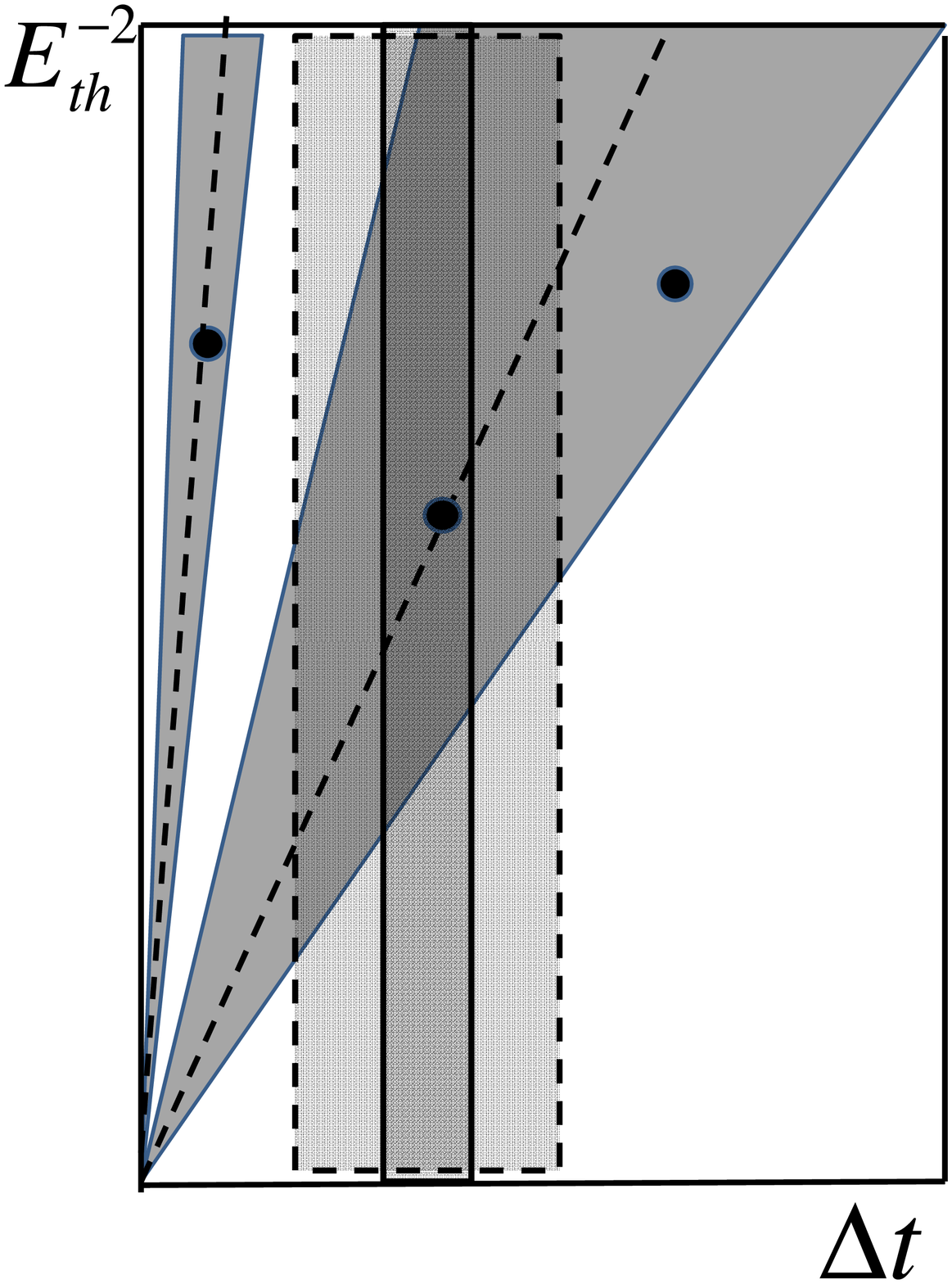} \\
\caption{\small On a plot of $1/E^2$ versus $t$ the figure shows the search windows used in the original 2007 analysis (solid and dashed rectangles), and the "two-triangle search window" that is based on the assumption of 4.0 eV and 21.4 eV mass eigenstates in SN 1987A.  Note that as the two-triangle search window slides along the $t$ axis, it would contain the 3 simulated events for the position shown. The value of the total horizontal width along the time axis $\Delta t$ of the two-triangle search window is proportional to $1/E_{th}^2$ and it is 182 s for $E_{th}=10 MeV,$ and 62s for $E_{th}=17 MeV.$    When $E_{th}=10 (17) MeV,$ the 20 s wide rectangular search window would be the skinny (fat) rectangle.   Hence, we see that both the fat and the skinny rectangles would contain only one event at a time (as it slides along $t$), and it would miss the hypothetical signal.  Moreover, the two-triangle search window with the choice $E_{th}=17 MeV$ would also miss the signal because in this case the 3 events would not be as depicted in the figure, but would be spread out in time three-fold from their depiction there, when the time axis is in units of seconds.
 \\}
\end{figure}

\end{document}